\documentstyle[prb,aps,epsf]{revtex}

\twocolumn

\begin{document}

\title{Hubbard-$U$ calculations for Cu from first-principle Wannier functions}
\author{I. Schnell and G. Czycholl}
\address{Department of Physics, University of Bremen, P.O.Box 330 440,
  D-28834 Bremen, Germany}
\author{R. C. Albers}
\address{Theoretical Division, Los Alamos National Laboratory, Los
  Alamos, New Mexico 87545}
\date{\today}
\maketitle
\begin{abstract}
We present first-principles calculations of optimally localized
Wannier functions for Cu and use these for an ab-initio determination
of Hubbard (Coulomb) matrix elements.  We use a standard linearized
muffin-tin orbital calculation in the atomic-sphere approximation
(LMTO-ASA) to calculate Bloch functions, and from these determine
maximally localized Wannier functions using a method proposed by
Marzari and Vanderbilt.  The resulting functions were highly
localized, with greater than 89\% of the norm of the function within
the central site for the occupied Wannier states.  Two methods for
calculating Coulomb matrix elements from Wannier functions are
presented and applied to fcc Cu.  For the unscreened on-site Hubbard
$U$ for the Cu 3d-bands we have obtained about 25eV.  These results
are also compared with results obtained from a constrained
local-density approximation (LDA) calculation.
\end{abstract}

\section{Introduction}
\label{sec:intro}

During the past few decades powerful numerical methods have been
developed for the ab-initio (first-principles) calculation of the
electronic ground-state properties of solids.  In most of these
methods density-functional theory\cite{HK64} (DFT) has been used to
treat the electron-electron Coulomb repulsion, and a local-density
approximation\cite{KS65} (LDA) (or a local spin-density approximation,
LSDA, for magnetic systems) has been used for the exchange-correlation
potential.  This procedure has been very successful for many materials
and ground-state properties (e.g., crystal structure, lattice
constant, binding energy, and ionization energy), but it has its
limitations; the band gap of semiconductors is not properly
reproduced, for instance.  Furthermore, for systems such as
high-temperature superconductors, heavy fermion materials,
transition-metal oxides, and 3d itinerant magnets, i.e., for systems
in which the Fermi level falls into a region of narrow energy bands,
the LDA is usually not sufficient.  It is generally accepted that the
problem for these materials is the strong electronic correlations that
are responsible for their electronic properties.  For a description of
such strongly correlated systems one usually instead uses as a
starting point model Hamiltonians like the Hubbard model\cite{H63} and
its multi-band generalizations.  But in these models the Coulomb
(interaction) matrix elements and also the one-particle (hopping)
matrix elements that determine the unperturbed band structure are
usually treated as free, adjustable parameters, i.e., they are not
known from ''first principles'' for the given material; on the other
hand, Coulomb correlations can be studied within reliable many-body
approximations that go beyond the Hartree-Fock approximation.

Both the ab-initio LDA and the many-body model-Hamiltonian methods
based on Hubbard-like models have their merits, but until rather
recently they have been almost separate and complementary approaches.
But, in view of the power of each, a combination of these methods is
desirable; and, in fact, during the last few years there have been
some attempts in this
direction.\cite{AZA91,SASS91,SAS9294,APKAK97,LK98,DJK99,KL99,LL00,ZPKPNA00,WPN00,NHBPAV00}
All of these recent developments add local, screened Coulomb (Hubbard)
correlations U between localized orbitals to the one-particle part of
the Hamiltonian obtained from an ab-initio LDA band-structure
calculation, but differ in how they handle the correlation part.  In
the earliest attempts, the LDA+U method\cite{AZA91} used essentially a
static mean-field-like (or Hubbard-I-like) approximation for the
correlation.  The simplest approximation beyond Hartree-Fock,
second-order perturbation theory (SOPT) in U, was
used\cite{SASS91,SAS9294,DJK99,WPN00} to study the electronic
properties of 3d-systems (like Fe and Ni) and heavy fermion systems
(like UPt$_3$).  The LDA++ approach\cite{LK98,KL99,LL00} has a similar
strategy, but uses other many-body approximations to treat the
correlation problem, namely, either the fluctuation exchange
approximation (FLEX) or the dynamical mean field theory\cite{GKKR96}
(DMFT).  Some of the other many-body
treatments\cite{APKAK97,ZPKPNA00,NHBPAV00} have also used DMFT, which
is based on the limit of large-dimension ($d \rightarrow \infty$)
approximation for correlated lattice electrons.\cite{MV89} Within DMFT
the selfenergy becomes local, i.e., independent of momentum $k$, which
allows a mapping of the lattice problem onto an effective impurity
model.  The LDA+DMFT
treatments\cite{KL99,LL00,ZPKPNA00,WPN00,NHBPAV00} mentioned above
differ in the many-body method they used for the effective impurity
problem, namely, Quantum Monte Carlo\cite{NHBPAV00} (QMC), the
non-crossing approximation\cite{ZPKPNA00} (NCA) or (iterated)
perturbation theory\cite{APKAK97} (IPT).  But all these approaches,
including the LDA+U, have in common that they have to introduce a
Hubbard U as an additional parameter, and hence are not real
first-principles (ab-initio) treatments.  Although they use an LDA
ab-initio method to obtain a realistic band structure, i.e.,
single-particle properties, Coulomb matrix elements for any particular
material are not known, and the Hubbard U remains an adjustable
parameter.

One can obtain estimates on the magnitude of $U$ either from
experiment (high-energy spectroscopy) or from the constrained LDA
method.\cite{AZA91,DBZA84,HSC88,MMS88,GAJZ89,MAM90,AAW93} Within the
latter method one adds the constraint that the electron occupation
number for the correlated bands is fixed to a given number through a
Lagrange parameter.  One can then use LDA to calculate the
ground-state energy for different occupations of the correlated
states, and the difference between the energy for double and single
occupation is an estimate for the Hubbard U.  This method has the
advantage that effects of screening are already somehow included.  On
the other hand there are usually several bands and many interaction
matrix elements (on-site, density-density, intra-band, inter-band,
exchange, inter-site, etc.) that have different magnitudes, and the
constrained LDA can only give some average value for these various
Coulomb matrix elements and not the individual ab-initio parameters
(Coulomb matrix elements).  This approach is intuitive and contains
some type of screening within a one-electron LDA approach.  It is
difficult to sort out the actual approximation involved.

In this paper we suggest a different approach, namely, the direct
ab-initio calculation of the one-particle (tight-binding) and
two-particle (Coulomb) matrix elements.  Our starting point is a
standard electronic band-structure calculation, for which we have used
the linearized muffin-tin orbital (LMTO) method\cite{HS84,OKA75}
within the atomic sphere approximation (ASA).  The LDA band-structure
calculation yields not only one-particle energies but also their
eigenstates, the Bloch wavefunctions, which form a proper basis of a
one-particle Hilbert space.  To determine the local (on- and
inter-site) Coulomb matrix elements it is necessary to construct
Wannier functions,\cite{GHW37,JC74} which are closely related to the
Bloch functions via a unitary transformation, but which are not 
unique since the phases of the Bloch functions are undetermined.

As first suggested by Marzari and Vanderbilt,\cite{MV97} this gauge
freedom can be used to construct ``maximally localized Wannier
functions.''  Those are just Wannier functions with a special gauge
that makes them optimally localized according to some criterion.  A
proper localization of the Wannier functions is important in our
opinion, because only then do the standard assumptions of the model
treatments hold such that only a few (on-site, nearest, and
next-nearest neighbor) one-particle (hopping) and two-particle
(Coulomb) matrix elements have to be considered explicitely.  These
matrix elements can then be calculated from the Wannier functions.

We use two different methods to calculate Coulomb matrix elements from
Wannier functions.  The first method uses the fact that the
LMTO-method provides Bloch functions in the basis of linear muffin-tin
orbitals.\cite{HS84} Therefore the Wannier functions are given as
linear combinations of such muffin-tin orbitals as well, and can be
used to evaluate the Coulomb integrals efficiently, similarly to what
was done in Ref.~\onlinecite{SA86}.  The second method uses a fast
Fourier transformation (FFT).  It does not rely on the property of the
wave functions being linear and is therefore more general.  It is also
very quick and efficient.

The paper is organized as follows.  In Sec.~\ref{sec:comp} we present
some of the computational details: we describe the form of the Bloch
functions in the LMTO method, how to obtain the Wannier functions from
them, and how to optimize the choice of the Wannier functions by using
the Marzari-Vanderbilt method.  Then we describe how the one-particle
(hopping) matrix elements are obtained from these localized Wannier
functions, and we present the two methods to calculate the Coulomb
matrix elements.  To illustrate the method we have performed actual
calculations for a well understood system, namely for Cu.  Although
this material is not a strongly correlated system, it has almost
completely filled, narrow, 3d-bands, for which (well localized)
Wannier functions and one- and two-particle matrix elements can be
calculated.  Results for Cu are presented in Sec.~\ref{sec:res}, where
we show some of the Wannier functions, demonstrate how well localized
they are and that the one-particle (tight-binding) matrix elements
obtained from them allow for a reconstruction of the band structure.
The direct Coulomb matrix elements obtained are rather large, between
20 and 25 eV for Wannier states with mainly 3d-character, and are
about 5 eV for nearest neighbors, and about 1 eV for exchange
interactions.  In Sec.~\ref{sec:constrain} we describe constrained LDA
calculations, which yield somewhat smaller values (about 18 eV) for
the Hubbard U of Cu, and in the final section (\ref{conc}) we discuss
how to extend and further apply the current approach.

\section{Computational Details}
\label{sec:comp}

We restrict ourself to the case where there is only one atom per unit
cell and where we can neglect spin (non-spin-polarized calculations).
For a given material the only input to an ab-initio LDA
calculation is the atomic number.  In the density-functional
approach,\cite{HK64,DG90} a local-density approximation is normally
used for the exchange and correlation interactions between the
electrons; we have used the von Barth-Hedin\cite{BH72}
exchange-correlation potential and a frozen-core approximation.
Within DFT the total energy of the ground state could be calculated as
a function of volume for a given crystal structure and used to
determine the equilibrium lattice constant; it is usually in good
agreement with experiment.  However, since our focus is on the
determination of Coulomb matrix elements in a Wannier basis, we have
simply used the experimental lattice parameters.

\subsection{LMTO wave functions}
\label{sec:lmto}

For our band-structure results we have used the LMTO
method\cite{HS84,OKA75} within the atomic sphere approximation (ASA).
The combined correction term\cite{HS84} was not included.  The
muffin-tin spheres are overlapping and their radius (the Wigner-Seitz
radius $S$) is determined by the condition that the sphere volume
equals the volume of the unit cell.  Within the muffin-tin spheres the
potential and wave functions are expanded in spherical harmonics with
a cutoff $l_{\rm max}=3$, i.e., s, p, d, and f-orbitals are included.
Furthermore the Bloch wave functions are given in terms of the
solution to the radial Schr{\"o}dinger equation $\phi_{\nu l}(r)$ to
some fixed energies $E_{\nu l}$ and its energy derivative $\dot
\phi_{\nu l}(r)$:
\begin{equation}
  \label{eq:lmto-wave}
  \Psi_{n {\bf k}} ({\bf r}) = \sum_L \left(
    \phi_{\nu l}(r) A_L^{n {\bf k}} + \dot \phi_{\nu l}(r) B_L^{n {\bf k}}
  \right)
  Y_L( \hat {\bf r} ) ~,
\end{equation}
where we use complex spherical harmonics in all of our calculations.
This expansion is valid in one muffin-tin sphere.  Here, as usual,
$L=\{l,m\}$ is understood and $n$ is the band index and ${\bf k}$ is
the wave vector. We define $n$ by the condition that $E_n({\bf
k})<E_{n+1}({\bf k})$.  The virtue of using this method for Wannier
functions is the simplification that only integrals over spheres are
needed; no real-space integrations over complicated Wigner-Seitz
(unit) cells are required.

The Bloch functions obey
\begin{equation}
  \label{eq:bloch}
  \Psi_{n {\bf k}} ({\bf r+R}) = e^{i{\bf k R}}
  \Psi_{n {\bf k}} ({\bf r}) ~.
\end{equation}
Therefore, the knowledge of a Bloch function in a single muffin-tin
sphere is sufficient for the knowledge of the function in the whole
crystal.  This situation is different when we consider Wannier
functions, which can be centered on different sites.  It is useful to
introduce a notation that holds for both Bloch and Wannier functions.
To do this we perform an expansion like Eq.~(\ref{eq:lmto-wave}) in
each muffin-tin sphere, which we label by its site vector ${\bf
R}$. The complete wave function (either Bloch or Wannier) is then
given by:
\begin{equation}
  \label{eq:asa}
  \Phi_\alpha ({\bf r}) = \sum_i \Phi_\alpha ({\bf R}_i; {\bf r}
   -{\bf R}_i)
\end{equation}
In this equation we have used the general notation for the
wave function expansion
$\Phi_\alpha ({\bf R}_i; {\bf r}-{\bf R}_i)$ such that:
(i) $\Phi$ is any kind of wave function.
(ii) The $\alpha$ stands for quantum numbers
(Bloch: $\alpha=\{ n, {\bf k} \}$;
Wannier: $\alpha=\{ {\bf R}, n \}$).
(iii) The first argument in the parenthesis indicates the muffin-tin
sphere about which we are expanding and is labeled by its site vector.
(iv) The second argument in the parenthesis is the position inside
this muffin-tin sphere described by its relative vector.  This means
that this vector has zero length in the center of the muffin-tin
sphere described by the first argument.
(v) Note that, for every ${\bf R}$,
\begin{equation}
  \nonumber 
  \Phi_\alpha ({\bf R}; {\bf r}) = 0 ~~~ {\rm if} ~ |{\bf r}|>S ~.
\end{equation}
In the case where $\Phi$ is a Bloch function we find
\begin{equation}
  \label{eq:basa}
  \Psi_{n{\bf k}} ({\bf R}; {\bf r}) = e^{i{\bf k R}}
  \Psi_{n{\bf k}}({\bf r}) ~.
\end{equation}
It is easy to see that Eq.~(\ref{eq:basa}) inserted in
Eq.~(\ref{eq:asa}) obeys Eq.~(\ref{eq:bloch}).

Also note that Eq.~(\ref{eq:asa}) disregards the effects of
overlapping muffin-tin spheres.  Within the ASA approximation, all
derivations are done as though non-overlapping muffin-tins are being
used, and then these formulas are used with expanded muffin-tins,
whose volumes sum to equal the unit cell volume (where the muffin-tin
radius is expanded to a Wigner-Seitz radius for one atom per unit
cell).  In addition, this approximate eliminates the necessity to
handle interstitial regions, and hence the ASA formalism is
mathematically much simpler than a full-potential electronic-structure
calculation would require.

\subsection{Wannier functions}
\label{sec:wan}

In this section we show how to calculate Wannier functions from the
LMTO type of Bloch functions of Eq.~(\ref{eq:lmto-wave}).  The Wannier
functions\cite{GHW37} are defined by
\begin{equation}
  \label{eq:wdef}
  w_{{\bf R}n}({\bf r}) \equiv \langle {\bf r} \vert {\bf R}n \rangle
  = \frac{1}{N} \sum_{\bf k} e^{-i{\bf kR}} \Psi_{n{\bf k}}({\bf r}) ~.
\end{equation}
Here, $N$ is the number of ${\bf k}$-mesh points in the Brillouin zone
or, equivalently, the number of unit cells in the real space supercell
that is used to discretize the ${\bf k}$-mesh.  As mentioned above,
Wannier functions are not unique.  Consider, for example, a single
band $n$ with Bloch functions $\vert \Psi_{n{\bf k}}\rangle$; a
transformation of the kind
\begin{equation}
  \label{eq:sbt}
  \vert \Psi_{n{\bf k}}\rangle ~~ \rightarrow ~~
  e^{i\phi_n^{({\bf k})}} \vert \Psi_{n{\bf k}}\rangle ~~,
  ~~~~~ \phi_n^{({\bf k})} \mbox{ real} ~,
\end{equation}
will still lead to Bloch functions. We shall call this a gauge
transformation of the first kind.  In the case of a composite set of
bands,\cite{MV97} this nonuniqueness corresponds to the freedom to
choose the phases and ``band-index labeling'' at each ${\bf k}$ point
of the Bloch functions:
\begin{equation}
  \label{eq:mbt}
  \vert \Psi_{n{\bf k}}\rangle ~~ \rightarrow ~~
  \sum_m U_{mn}^{({\bf k})}\,\vert \Psi_{m{\bf k}}\rangle
\end{equation}
We shall call this a gauge transformation of the second kind.  Here
$U_{mn}^{({\bf k})}$ is a unitary matrix.  From all the arbitrary
choices of Wannier functions we will pick out that particular set that
minimizes the total spread given by
\begin{equation}
  \label{eq:om}
  \Omega = \sum_n \left[ \langle r^2\rangle_n -
                         \langle r \rangle^2_n \right] ~.
\end{equation}
For any operator $A$, $\langle A \rangle_n$ denotes the expectation
value $\langle {\bf R} n \vert A \vert {\bf R} n \rangle$.  A method
for minimizing Eq.~(\ref{eq:om}) has been developed by Marzari and
Vanderbilt\cite{MV97} and its application to the ASA wave functions
does not pose any particular problems (details will be given below).

Before minimizing $\Omega$ according to this procedure, it is useful
to prepare the Bloch orbitals to make the starting Wannier functions
somewhat localized.  This has two advantages: (i) the minimization
procedure converges faster, and (ii) this helps avoid getting trapped
in local minima.  Marzari and Vanderbilt\cite{MV97} suggest several
possible preparations.  We have found our own method, which seems to
work well.  This involves a simple gauge transformation for each band,
which is given by
\begin{equation}
  \label{eq:il}
  \Psi_{n{\bf k}}({\bf r}) \rightarrow
  \exp(-i{\rm Im}\,\ln \Psi_{n{\bf k}}({\bf r}_0))
  \Psi_{n{\bf k}}({\bf r}) ~.
\end{equation}
This gauge transformation has the property that ${\rm Im}\,\ln
\Psi_{n{\bf k}}({\bf r}_0)$ transforms to zero.  So at the point ${\bf
r}_0$ all the Bloch functions will have the same phase (in this case
just $1+i0$) and $\langle {\bf r}_0 \vert {\bf 0}n \rangle$ will take
a large value.  We thus expect the Wannier function to be fairly
localized at ${\bf r}_0$.  To make the method work well one should
choose ${\bf r}_0$ where the Wannier functions are expected to be
reasonably large.  In our calculations we have chosen the direction of
this vector to be well away from the expected zeroes of the spherical
harmonics and with an absolute value far enough away from the origin
to be in a place where the Wannier functions should have a significant
magnitude.  We found an ${\bf r}_0$ of (0.8, 1.0, 0.3)$a_0$ to work
well for fcc Cu.

We shall now derive expressions of the form of Eq.~(\ref{eq:asa}) for
Wannier functions.  From Eqs.~(\ref{eq:basa}) and (\ref{eq:wdef}) we
have
\begin{eqnarray}
  \label{eq:wcent}
  w_{{\bf R}n} ({\bf R'}; {\bf r}) &=&
  \frac{1}{N} \sum_{\bf k} e^{-i{\bf kR}} \Psi_{n{\bf k}} ({\bf R'}; {\bf r})
\nonumber \\
  &=& \frac{1}{N} \sum_{\bf k}
      e^{i{\bf k}({\bf R'-R})} \Psi_{n{\bf k}}({\bf r}) ~.
\end{eqnarray}
Because Wannier functions on different sites have the same form
(shape) of their wave functions and differ only by a translation of
their origin, it is useful to use a notation that indicates values of
a wave function relative to a Wannier function centered at the origin:
\begin{equation}
  \label{eq:wtrs}
  w_{{\bf R}n} ({\bf R'}; {\bf r}) =
  w_{{\bf 0}n} ({\bf R'-R}; {\bf r}) \equiv
  w_n ({\bf R'-R}; {\bf r}) ~,
\end{equation}
where we have introduced the notation $w_{{\bf 0}n} \equiv w_n$ (i.e.,
if the subscript contains only a wave function label without a spatial
vector ${\bf R}$, then we are using a relative notation that refers to a
Wannier function centered at the origin).  We can use the Bloch
condition (cf.~Eq.~(\ref{eq:wcent})) to calculate the parts of the
Wannier function on other sites ${\bf R}$:
\begin{equation}
  \label{eq:wasa}
  w_n ({\bf R}; {\bf r}) =
  \frac{1}{N} \sum_{\bf k} e^{i{\bf kR}} \Psi_{n{\bf k}}({\bf r}) ~.
\end{equation}
Note that $|{\bf r}|<S$.  For this notation to work in our numerical
calculations, it is essential to force the Wannier center, i.e., the
muffin-tin sphere where $\langle {\bf r} \vert {\bf 0}n \rangle$ is
largest, to be at the muffin-tin sphere around the lattice site ${\bf
0}$; we achieve this by setting $|{\bf r}_0|<S$ in Eq.~(\ref{eq:il}).
In most of the rest of the paper, we will almost always use the
relative notation that refers to Wannier states centered at the
origin, and will perform whatever translations are necessary to be
able to use these states.

In the method of Marzari and Vanderbilt,\cite{MV97} the starting point
for the calculations are a set of reference matrices defined by
\begin{equation}
  \label{eq:refM}
  M^{(0)({\bf k,b})}_{mn} = \langle \Psi_{m{\bf k}} 
    \vert e^{-i{\bf br}} \vert \Psi_{n,{\bf k+b}} \rangle ~.
\end{equation}
Here ${\bf b}$ denotes a nearest-neighbor vector on the discretized
mesh in k-space (in this method, the set of ${\bf b}$-vectors are
needed for numerical derivatives).  We calculated the action of
$e^{-i{\bf br}}$ on the ket by using Eqs.~(\ref{eq:SH3}) and
(\ref{eq:SH2}) and solved the remaining integral by using
Eq.~(\ref{eq:I3}).  We used a uniform (cubic) discrete ${\bf k}$-mesh
with a spacing $\Delta k$ of $0.2 (2\pi/a)$.  In such a mesh there are
6 nearest-neighbors for the ${\bf b}$ vectors needed for the numerical
derivatives.  We were careful not to double count vectors in the ${\bf
k}$-mesh (those equivalent to each other by a reciprocal lattice
vector) within the Brillouin zone (which has 500 ${\bf k}$ points in
the full zone).

We then used the steepest-descent method and relevant equations in
Sec.~IV of Ref.~\onlinecite{MV97} to iterate a series of small steps
where a set of $\Delta W^{\bf k}$ were calculated and used to update
the unitary matrices $U^{\bf k}$ and the $M^{\bf k,b}$ matrices.
After each iteration, where we update all the relevant ${\bf k}$
matrices, we calculated the spread function $\Omega$, and continued
iterating until this converged.

In these calculations the initial matrices $M^{(0)({\bf k,b})}$ are by
far the most time consuming computationally (it requires storing
$6\cdot500\cdot16^2=768,000$ complex numbers).  The iterations of the
steepest descent method were much faster.  For this reason we used
many iterations (about 1500 steps) and converged $\Omega$ to about
0.01\%.  For the step size (cf.\ Eq.~(57) of Ref.~\onlinecite{MV97})
we used an $\alpha$ of 0.2.

The final result can be written in a form similar to the LMTO wave
functions, Eq.~(\ref{eq:lmto-wave}),
\begin{equation}
  \label{eq:wann-wave}
  w_n({\bf R;r}) = \sum_L \left(
    \phi_{\nu l}(r) A_L^{n {\bf R}} + \dot \phi_{\nu l}(r) B_L^{n {\bf R}}
  \right)
  Y_L( \hat {\bf r} ) ~,
\end{equation}
where the $A$ and $B$ matrices originally come from the LMTO wave
functions, but are then updated from the relevant phase information,
unitary matrix, and other integrations and transformations of the
method.

Because of the normalization of the starting LMTO Bloch wave functions
(which are normalized to unity within a single unit cell), each
Wannier function is naturally normalized to unity when integrated over
all space.

\subsection{One particle matrix elements}
\label{sec:one}

The Wannier function basis can be viewed as an orthogonal
tight-binding basis.  For this reason it is useful to calculate
one-particle matrix elements of the Hamiltonian in the Wannier basis.
As we shall see, these matrix elements are (for a gauge transformation
of the first kind only) equivalent to the Fourier components of the
band structure; this equivalence is useful for checking some of the
numerical aspects of the calculations.

Because the Hamiltonian has the property that $H({\bf r})=H({\bf r+R})$,
it is sufficient to consider the matrix elements:
\begin{equation}
  \label{eq:tRnm}
  t_{{\bf R}nm} \equiv \langle {\bf R}n \vert H \vert {\bf 0}m \rangle
\end{equation}
Inserting Eq.~(\ref{eq:wdef}) and using
$H|\Psi_{n\bf k}\rangle = E_n({\bf k})|\Psi_{n\bf k}\rangle$ we find
\begin{equation}
  \label{eq:tEnk}
  t_{{\bf R}nm} = \frac{\delta_{nm}}{N}
  \sum_{\bf k} e^{i{\bf kR}} E_n({\bf k}) ~,
\end{equation}
which are just the Fourier components of the band structure.  The
Bloch states $|\Psi_{n\bf k}\rangle$ continue to be eigenstates of $H$
under a gauge transformation of the first kind and one can also easily
show that the $t_{{\bf R}nm}$ are invariant under this type of gauge
transformation.  The $t_{{\bf R}nm}$ from Eq.~(\ref{eq:tEnk}) can be
directly calculated from the band structure $E_n({\bf k})$.

A gauge transformation of the second kind leads to states $|\Psi_{n\bf
k} \rangle$ that are no longer eigenstates of $H$ with eigenvalue
$E_n({\bf k})$.  Therefore $t_{{\bf R}nm}$ are not invariant under a
gauge transformation of the second kind. However, generally we can
always calculate
\begin{equation}
  \label{eq:Hk}
  H^{\bf k}_{nm} \equiv \langle \Psi_{n{\bf k}} |
  H | \Psi_{m{\bf k}} \rangle =
  \sum_{\bf R} e^{-i{\bf kR}} t_{{\bf R}nm}
\end{equation}
and use its diagonalized eigenvalues as the band structure for any set
of Wannier functions.  The matrix $H^{\bf k}_{nm}$ is Hermitian; it
is, of course, already diagonal for a gauge transformation of the
first kind, with the energy levels as the diagonal matrix elements.

To calculate $t_{{\bf R}nm}$ from Eq.~(\ref{eq:tRnm}), i.e., using
Wannier functions, we can use Eqs.~(\ref{eq:asa}) and (\ref{eq:wtrs})
to find
\begin{equation}
  \label{eq:tasa}
  t_{{\bf R}nm} = \sum_i \int d^3{\bf r} ~
  w^*_n ({\bf R}_i-{\bf R}; {\bf r}) H w_m ({\bf R}_i; {\bf r}) ~,
\end{equation}
where the integral is over a single sphere only.  The effect of $H$ on
the second wave function can be carried out easily because we are
working in a linear basis. We only note that $(H-E_{\nu l})
\phi_l(r)=0$ and $(H-E_{\nu l}) \dot \phi_l(r) = \phi_l(r)$, for
details see Ref.~\onlinecite{HS84}. In order to calculate
Eq.~(\ref{eq:tasa}), we must evaluate integrals of the form
\begin{equation}
  \label{eq:I1}
  I = \int d^3{\bf r} f^*_1({\bf r}) f_2({\bf r}) ~,
\end{equation}
where the functions $f_i({\bf r})$ are given by the expansion
\begin{equation}
  \label{eq:I2}
  f_i({\bf r}) = \sum_L R_{iL}(r) Y_L(\hat{\bf r}) ~.
\end{equation}
Inserting Eq.~(\ref{eq:I2}) into Eq.~(\ref{eq:I1}) and using the
orthonormality of the spherical harmonics yields:
\begin{equation}
  \label{eq:I3}
  I = \sum_L \int dr ~ r^2 ~ R^*_{1L}(r) ~ R_{2L}(r)
\end{equation}
Because Eq.~(\ref{eq:lmto-wave}) was our starting point, the radial
functions $R(r)$ will always be given in terms of $\phi_l(r)$ and
$\dot\phi_l(r)$, i.e.,
\begin{equation}
  \label{eq:I4}
  R_{iL}(r) = A_{iL}\phi_l(r) + B_{iL}\dot\phi_l(r) ~.
\end{equation}
We will use this form to calculate the integral $I$ very
efficiently. It is clear that any integral can be reduced to a linear
combination of ``basic'' integrals. Those basic integrals consist of
the (very limited) combinations of the $\phi_l(r)$'s and
$\dot\phi_l(r)$'s. We will label them by
\begin{eqnarray}
  \label{eq:I5}
  b_{l;p_1p_2} = && \int dr \; r^2 \;
  [ \delta_{p_1,0} \phi_l(r) + \delta_{p_1,1} \dot \phi_l(r) ] \nonumber \\
  && \times [ \delta_{p_2,0} \phi_l(r) + \delta_{p_2,1} \dot \phi_l(r) ] ~,
\end{eqnarray}
where $p_1$ and $p_2$ can take the values 0 and 1. So it must be
possible to write the integral $I$ as:
\begin{equation}
  \label{eq:I6}
  I = \sum_L \sum_{p_1=0}^1 \sum_{p_2=0}^1 a_{L;p_1p_2} b_{l;p_1p_2}
\end{equation}
It follows that the coefficients $a_{L;p_1p_2}$ are given by
\begin{eqnarray}
  \label{eq:I7}
  a_{L;p_1p_2} = && 
  [ \delta_{p_1,0} A^*_{1L} + \delta_{p_1,1} B^*_{1L} ] \nonumber \\
  && \times [ \delta_{p_2,0} A_{2L} + \delta_{p_2,1} B_{2L} ] ~.
\end{eqnarray}
We are now in a position to calculate Eq.~(\ref{eq:tasa}) with the aid
of Eq.~(\ref{eq:I6}).

\subsection{Wannier-Function Projected Density of States}

The density of states (DOS) per spin is defined by
\begin{equation}
  \label{eq:defDOS}
  N(E)=\frac{V}{(2\pi)^3} \sum_n \int_{BZ}
  d^3{\bf k}~\delta(E-E_n({\bf k})) ~,
\end{equation}
where $V$ is the volume of the unit cell. In the same way that the DOS
is often projected in terms of the $l$-character of the states, one
can do a similar treatment for a projection onto the Wannier states.
We can define a projected DOS for Wannier states, by inserting the
projection operator onto the Wannier states $|{\bf 0 }j \rangle
\langle {\bf 0}j|$ into Eq.~(\ref{eq:defDOS}):
\begin{equation}
  \label{eq:wDOS} N_j(E) = \frac{V}{(2\pi)^3} \sum_n \int_{BZ}
  d^3{\bf k}~ |\langle
  \psi_{n{\bf k}} | {\bf 0}j \rangle|^2 \delta(E-E_n({\bf k}))
\end{equation}
Note that the $\psi_{n{\bf k}}$ in this formula have to be the Bloch
states before the gauge transformation, since the band structure
$E_n({\bf k})$ is related to the untransformed states.  The Bloch wave
functions are normalized to a single unit cell and each Wannier
function over all space.  We can calculate $N_j(E)$ by using the
tetrahedron method.\cite{Tetra71} For the ${\bf k}$-points that form
the tetrahedras we need to calculate $|\langle \psi_{n{\bf k}} | {\bf
0}j \rangle|^2$, which we have done using the scheme described in
section \ref{sec:one}.  In these calculations it is important to be
aware that $ | {\bf 0}j \rangle$ has parts of its wavefunction on
other sites than the central site where it is centered.  In our
calculations, we included parts of the Wannier function out to 17
near-neighbor shells of sites.

Note that the exact projection operator is a sum over all ${\bf R}$,
since
\begin{equation}
  \sum_{{\bf R}j} | {\bf R}j \rangle \langle {\bf R}j | = 1 ~.
\end{equation}
However, it is sufficient to only consider the Wannier states $|{\bf
0}j\rangle$ in our projection (and not all the $|{\bf R}j\rangle$),
since
\begin{equation} \nonumber
  |\langle \psi_{n{\bf k}} | {\bf R}j \rangle| =
  |\langle \psi_{n{\bf k}} | {\bf 0}j \rangle| ~.
\end{equation}
We can check the correctness of our projection by comparing
$N_{tot}(E)=\sum_j N_j(E)$ with the $N(E)$ that is calculated directly
from the LMTO energy eigenvalues.  We find that our projected sum is
accurate to within 0.2\% of the LMTO value.

\subsection{Coulomb matrix elements}
\label{sec:coul}

The matrix elements we wish to calculate are
\begin{equation}
  \label{eq:W1234}
  W_{12,34} = \int \frac{d^3{\bf r} d^3{\bf r'} ~ e^2}{|{\bf r}-{\bf r'}|}
  w^*_1({\bf r}) w^*_2({\bf r'}) w_3({\bf r}) w_4({\bf r'}) ~.
\end{equation}
where $1,2,3,4=i=\{{\bf R}_in_i\}$ is a Wannier state, and $W$ denotes the
Coulomb interaction.  The spatial integrals over ${\bf r}$ and ${\bf r'}$
extend over all space. Using Eqs.~(\ref{eq:asa}) and (\ref{eq:wtrs}), we can
use translations to rewrite this expression so that the integrals are only
over the muffin-tin sphere at the origin:
\begin{equation}
  \label{eq:W1234a}
  W_{12,34} = \sum_{i,j} W(12,34;{\bf R}_i,{\bf R}_j) ~,
\end{equation}
where the expression $W(12,34;{\bf R},{\bf R'})$ is defined by:
\begin{eqnarray}
  \label{eq:W1234b}
  && \int \frac{d^3{\bf r} d^3{\bf r'}~~~e^2}
  {|{\bf r}-{\bf r'}+{\bf R}-{\bf R'}|}
  w^*_{n_1}({\bf R}-{\bf R}_1;{\bf r})
  w^*_{n_2}({\bf R'}-{\bf R}_2;{\bf r'})
  \nonumber \\
  && ~~~~~~~~~ \times
  w_{n_3}({\bf R}-{\bf R}_3;{\bf r})
  w_{n_4}({\bf R'}-{\bf R}_4;{\bf r'})
\end{eqnarray}

Since most applications of the Hubbard model use only two orbitals
instead of all four, it is useful to define the limiting subset of the
$W$ functions as direct Coulomb $U_{ij}$ and exchange $J_{ij}$
integrals:
\begin{eqnarray}
  \label{eq:UJ}
  U_{12} && = W_{12,12} \nonumber \\
  J_{12} && = W_{12,21} ~,
\end{eqnarray}
and the obvious generalizations for:
\begin{eqnarray}
  \label{eq:UJR}
  U(12;{\bf R, R'}) && = W(12,12;{\bf R, R'}) \nonumber \\
  J(12;{\bf R, R'}) && = W(12,21;{\bf R, R'})
\end{eqnarray}

\subsubsection{Spherical harmonics expansion}
\label{sec:met1}

We will now only consider matrix elements between Wannier functions
centered on the origin (i.e., where the ${\bf R}_i={\bf 0}$ in
Eq.~(\ref{eq:W1234}).  Because we are using maximally localized
Wannier functions, most of the Wannier functions have their largest
component in the center cell (see section \ref{sec:res}).  As a first
approximation, we will therefore neglect all other muffin-tin spheres.
This approximation allows us to calculate on-site inter-band matrix
elements.  We are thus looking for
\begin{eqnarray}
  \label{eq:I10}
  W_{n_1n_2,n_3n_4} && \approx W(12,34;{\bf 0},{\bf 0}) \nonumber \\
  && = \int d^3{\bf r} d^3{\bf r'}
  w^*_{n_1}({\bf 0};{\bf r}) w^*_{n_2}({\bf 0};{\bf r'}) \nonumber \\
  && \times \frac{ e^2 }{|{\bf r}-{\bf r'}|}
  w_{n_3}({\bf 0};{\bf r}) w_{n_4}({\bf 0};{\bf r'}) ~,
\end{eqnarray}
where the integral over ${\bf r}$ is only over the central site.
Inserting the expansion Eq.~(\ref{eq:I2}) for the Wannier functions
and making use of the well-known expansion (see for example
Ref.~\onlinecite{JDJ75})
\begin{equation}
  \label{eq:I11}
  \frac{1}{|{\bf r}-{\bf r'}|} = \sum_{\l=0}^{\infty}
  \frac{4\pi}{2l+1} \; \frac{r_<^l}{r_>^{l+1}}\sum_{m=-l}^l
  Y_L^*(\hat {\bf r'}) Y_L(\hat {\bf r}) ,
\end{equation}
where $r_>$ ($r_<$) is the length of the greater (smaller) of the two
vectors ${\bf r}$ and ${\bf r'}$, we find
\begin{eqnarray}
  \label{eq:I12}
  I = && \sum_l \frac{4\pi}{2l+1} \sum_{L_i}
  \int dr r^2 R_{1L_1}^*(r) R_{3L_3}(r) \int dr' {r'}^2 \times
\nonumber \\
  && \times  R_{2L_2}^*(r') R_{4L_4}(r') \frac{r_<^l}{r_>^{l+1}}
  \sum_{m=-l}^l C_{L_3L_1L} ~ C_{L_2L_4L} ~.
\end{eqnarray}
The coefficients $C_{LL'L''}$ are called Gaunt coefficients [see
Eq.~(\ref{eq:gaunt}) in the appendix].  If we define
\begin{equation}
  \label{eq:I13}
  C_{l;L_1L_2L_3L_4} \equiv \sum_{m=-l}^l C_{L_3L_1L} ~ C_{L_2L_4L}
\end{equation}
and
\begin{eqnarray}
  \label{eq:I14}
  I_{l;L_1L_2L_3L_4} \equiv && \int dr r^2 R_{1L_1}^*(r) R_{3L_3}(r)
\nonumber \\
  && \times \int dr' {r'}^2 R_{2L_2}^*(r') R_{4L_4}(r')
  \frac{r_<^l}{r_>^{l+1}} ~,
\end{eqnarray}
the integral takes the form
\begin{eqnarray}
  \label{eq:I15}
  I = \sum_{l,L_i} \frac{4\pi}{2l+1} C_{l;L_1L_2L_3L_4} I_{l;L_1L_2L_3L_4} ~.
\end{eqnarray}
The task is now to determine $I_{l;L_i}$ (we use the shorthand
notation $L_i$ for $L_1L_2L_3L_4$). To do this, we will use the
formalism developed in the last section. In complete analogy to
Eqs.~(\ref{eq:I5}-\ref{eq:I7}) we now find
\begin{equation}
  \label{eq:I16}
  I_{l;L_i} = \sum_{p_i} a_{L_ip_i} ~ b_{l;l_ip_i} ~,
\end{equation}
where
\begin{eqnarray}
  \label{eq:I17}
  a_{L_ip_i} = &&
  \prod_{i=1}^2 [ \delta_{p_i,0} A_{iL_i}^* +
  \delta_{p_i,1} B_{iL_i}^* ]
\nonumber \\
  && \times \prod_{i=3}^4 [ \delta_{p_i,0} A_{iL_i} +
  \delta_{p_i,1} B_{iL_i} ]
\end{eqnarray}
and
\begin{eqnarray}
  \label{eq:I18}
  b_{l;l_ip_i} = && \int dr ~ r^2
  [ \delta_{p_1,0} \phi_{l_1}(r) + \delta_{p_1,1} \dot \phi_{l_1}(r) ]
\nonumber \\
  && \times  [ \delta_{p_3,0} \phi_{l_3}(r)
             + \delta_{p_3,1} \dot \phi_{l_3}(r) ]
\nonumber \\
  && \times \int dr' ~ {r'}^2
     [ \delta_{p_2,0} \phi_{l_2}(r') +
     \delta_{p_2,1} \dot \phi_{l_2}(r') ]
\nonumber \\
  && \times [ \delta_{p_4,0} \phi_{l_4}(r')
            + \delta_{p_4,1} \dot \phi_{l_4}(r') ]
  \frac{r_<^l}{r_>^{l+1}} ~.
\end{eqnarray}
It should be noted that these basic integrals are symmetric with
respect to some of their indices. If we introduce the joined index
$n_i=\{\l_i,p_i\}$ then:
\begin{eqnarray}
  \label{eq:I19}
  b_{l;n_1n_2n_3n_4} = && b_{l;n_3n_2n_1n_4} = b_{l;n_1n_4n_3n_2} =
\nonumber \\
  b_{l;n_3n_4n_1n_2} = && b_{l;n_2n_1n_4n_3} = b_{l;n_2n_3n_4n_1} =
\nonumber \\
  b_{l;n_4n_1n_2n_3} = && b_{l;n_4n_3n_2n_1}
\end{eqnarray}
If we consider the numerical aspects for the case where $s$, $p$, $d$,
and $f$ orbitals are included in the wave function expansion, we find
that we need to use a cutoff of $l_{\rm max}=6$ in Eqs.~(\ref{eq:I11})
and (\ref{eq:I13}).  Using the symmetries in Eq.~(\ref{eq:I19}), we
then find 9072 basic integrals that have to be calculated and stored.
The sum in Eq.~(\ref{eq:I15}) however involves $7\cdot16^4=458,752$
elements. Fortunately, only 6778 combinations of the
$l,L_1,L_2,L_3,L_4$ coefficients in Eq.~(\ref{eq:I16}) have to be
calculated; the others vanish.  Each of these coefficients involves a
sum over 16 elements, and each of these elements is a product of 5
numbers.

\subsubsection{Fast Fourier Transformation (FFT) approach}
\label{sec:fft}

The method we have just described works well, but requires a lot of
Gaunt functions and other complications.  As written, it also only
involves integrals over the central site and ignores parts of the
Wannier functions on nearby neighbors.  We have therefore found a
different approach to the problem.

To calculate $W(12,34;{\bf R},{\bf R'})$ for any lattice sites ${\bf
R}$ and ${\bf R'}$, we make use of the Fourier transform
\begin{equation}
  \label{eq:ft1r}
  \int d^3{\bf q} ~ \frac{e^{i{\bf qr}}}{{\bf q}^2} =
  \frac{2\pi^2}{|{\bf r}|}
\end{equation}
and find for Eq.~(\ref{eq:W1234b}):
\begin{eqnarray}
  \nonumber
&& W(12,34;{\bf R},{\bf R'}) = \frac{e^2}{2\pi^2}
   \int \frac{d^3{\bf q}}{{\bf q}^2}
  ~ e^{i{\bf q}({\bf R}-{\bf R'})} f_{13}({\bf q}) f_{24}({\bf -q}) \\
  \label{eq:WM2}
&& f_{ij}({\bf q}) \equiv \int d^3{\bf r} e^{i{\bf qr}}
  w^*_{n_i}({\bf R}-{\bf R}_i;{\bf r})
  w_{n_j}({\bf R}-{\bf R}_j;{\bf r})
\end{eqnarray}
The $f_{ij}$ functions are just the Fourier transforms of a product of
some Wannier functions in a sphere. These can be calculated very
efficiently by calculating the Wannier functions on a cubic mesh in
real space and then applying a standard FFT algorithm.  To do this, we
have used the routine ``fourn'' (cf.~Ref.~\onlinecite{NRC92}).  For
details on how to apply the FFT to continuous functions,
Ref.~\onlinecite{FFT74} is very useful.  The result of the Fourier
transform is $f_{ij}({\bf q})$ on a cubic mesh in ${\bf q}$-space with
some $\Delta q$ (the distance between the mesh points). We perform the
remaining ${\bf q}$-integral in the following way. Let us call the
integrand without the $q^{-2}$ term
\begin{equation}
  \label{eq:Fq}
  F({\bf q}) = e^{i{\bf q}({\bf R}-{\bf R'})}
  f_{13}({\bf q}) f_{24}({\bf -q}) ~,
\end{equation}
which is smooth function at ${\bf q}={\bf 0}$. In order to treat the
divergence arising from $q^{-2}$, we split the integral in the following way:
\begin{equation}
  \label{eq:intFq}
  \int d^3{\bf q} \frac{F({\bf q})}{q^2} =
  \int d^3{\bf q} \frac{F({\bf q})-F({\bf 0})}{q^2} +
  F({\bf 0}) \int \frac{d^3{\bf q}}{q^2}
\end{equation}
All integrals are over a cube with length $N\Delta q$. The last integral is
just half of this length times $C$, which we define as
\begin{equation}
  \label{eq:mqw}
  C \equiv \int_{-1}^{+1}\!dx \int_{-1}^{+1}\!dy
  \int_{-1}^{+1}\!dz~\frac{1}{r^2} \approx 15.34825 ~.
\end{equation}
The remaining integral in Eq.~(\ref{eq:intFq}) is transformed into a
sum over little cubes with volume $(\Delta q)^3$. For ${\bf q}={\bf
0}$ the value of integrand is calculated via the second derivative of
$F({\bf q})$ at ${\bf q}={\bf 0}$ numerically (the second derivative
is needed to cancel the $q^2$ in a power-law expansion of $F$).

The cubic grid in real space that we used to calculate the Wannier
functions in Eq.~(\ref{eq:WM2}) had $N=64^3$ points in the real space
grid with a spacing $\Delta x=0.17$. The $\Delta q$ spacing of the
$q$-mesh is determined by $N$ and $\Delta x$.  Using the FFT for
continuous Fourier transformations one has to be very careful about
the choice of these values because the FFT is a discrete Fourier
transform.  It is important to make sure that the results of a FFT
calculation do not depend on the values $N$ and $\Delta x$.

Note that each integral in Eq.~(\ref{eq:WM2}) could be calculated from
the spherical-harmonic expansions.  However, many such integrals would
be required and the method would be extremely computationally
expensive.  The FFT method generates all the ${\bf q}$ values needed
with a single calculation and is much more efficient.  However,
because of finite mesh sizes and compromises between real and $q$
space integrals, it is not as accurate as the spherical-harmonic
expansion method of Sec.~\ref{sec:met1}, when the latter is applicable.

\section{Results and Discussion}
\label{sec:res}

We have tested our methods on Cu, which has the following properties:
(i) It has a simple close-packed fcc crystal structure for which the
ASA should be a reasonable approximation. (ii) Cu is a simple metal
that belongs to the 3$d$ transition metals, so one can determine
Coulomb matrix elements for the 3$d$ states which are interesting and
of relevance for the really correlated 3$d$-systems. (iii) Since Cu is
non-magnetic, we do not have to worry about spin-polarized or magnetic
calculations.  We have used the experimental lattice constant
$a$=3.614{\AA} as given in Ref.~\onlinecite{AM76}.  As usual we use
atomic Rydberg units and $a_0=\hbar^2/me^2$ is the Bohr radius.

\begin{figure}
\epsfxsize=3.0 truein
\centerline{\epsfbox{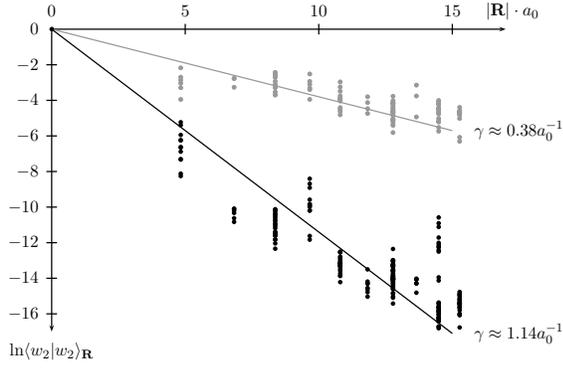}}
\caption{Localization of Wannier functions before (grey) and after (solid)
  minimization of the spread functional $\Omega$.  Each dot represents the
  portion of the wave function in a muffin-tin sphere. The best
  exponential fit to the decay is roughly $e^{-\gamma r}$ with
  the values of $\gamma$ given within the figure.
  In Eq.~(\ref{eq:il}) we set ${\bf r}_0=(0.8,1.0,0.3)a_0$.}
\label{fig:ls}
\end{figure}

From the one-electron Bloch wavefunctions, the Wannier functions were
obtained using 500 ${\bf k}$-points in the BZ. Since we have used a
cutoff of $l_{\rm max}=3$, the LMTO method generates 16 bands. In the
minimization procedure all 16 bands were treated as a composite set of
bands.  To demonstrate the localization of the Wannier functions
obtained, we have calculated
\begin{equation}
\langle w_n|w_n \rangle_{\bf R} = \int d^3r |w_n({\bf R};{\bf r})|^2 ~,
\end{equation}
which is the relative weight the Wannier function localized at the
site {\bf 0} has in the muffin-tin sphere centered around ${\bf R}$. In
Fig.~\ref{fig:ls} we have plotted for $n=2$ the function $\ln
\langle w_2|w_2 \rangle_{\bf R}$ as a function of $|{\bf R}|$. Although
the contribution to $w_2$ appears to decrease exponentially with
increasing distance from the central sphere when plotted in this way,
i.e., our Wannier functions are exponentially localized, we actually
get just as good a fit through the scatter of the data with a
power-law dependence with a power of about -7.  It is not easy to
numerically decide whether the decay is an exponential or a power-law
dependence, since our Wannier functions are ultimately periodic in a
supercell determined by the $\Delta \bf k$ spacing of the discrete
${\bf k}$ mesh used to construct them.  In either case, the Wannier
function is highly localized.  The grey dots and line in the figure
show the result when only the phase has been adjusted according to
Eq.~(\ref{eq:il}); then the Wannier functions have a relatively
smaller decay constant of $\gamma = 0.38 a_0^{-1}$.  The black dots
and line show the result after the full localization (minimization)
procedure of Ref.\onlinecite{MV97} has been applied by minimizing the
full set of all 16 bands considered; clearly a much better
localization with a larger decay constant $\gamma = 1.14 a_0^{-1}$ has
been achieved.  When we tried to minimize a smaller subset of bands (5
bands instead of the full 16) the decay factor was in between the
other two values, with $\gamma \approx .71a_0^{-1}$ (not shown in the
figure).
\parbox{7cm}{
\begin{table}
  \caption{Angular character of Wannier functions in the center muffin-tin
    sphere (top) and first shell, i.e., 12 nearest neighbors (bottom).}
  \label{tab:lcomp}
  \begin{tabular}{c|ccccccc}
  l    & $n=0$ &    1 &    2 &    3 &    4 &    5 &    6 \\ \hline
  (4)s &  .102 & .013 & .000 & .008 & .026 & .134 & .187 \\
  (4)p &  .297 & .131 & .058 & .042 & .151 & .373 & .392 \\
  (3)d &  .407 & .663 & .895 & .886 & .716 & .323 & .256 \\
  (4)f &  .096 & .140 & .024 & .039 & .064 & .069 & .051 \\
$\sum$ &  .902 & .947 & .977 & .975 & .956 & .899 & .886 \\ \hline
  (4)s &  .009 & .004 & .002 & .002 & .004 & .009 & .011 \\
  (4)p &  .019 & .008 & .004 & .004 & .008 & .020 & .023 \\
  (3)d &  .052 & .029 & .012 & .013 & .024 & .053 & .060 \\
  (4)f &  .012 & .008 & .003 & .003 & .006 & .012 & .013 \\
$\sum$ &  .092 & .049 & .021 & .023 & .041 & .094 & .107 \\
  \end{tabular}
\end{table}}
Here we should note that the Wannier functions are not pure in terms
of their $l$-character.  Table \ref{tab:lcomp} shows the angular
character in the center muffin-tin (MT) and the first shell for the
first 7 Wannier states.  We see that for the states with $n$=0 to $n$=4 the
 $d$-character is largest which suggests to call these states d-like
states yielding 5 d-states per spin direction as expected. But among these
states the d-character is highest (nearly 90 \%) for the
states $n$ = 2 and 3.  Table \ref{tab:lcomp} also tells us how much 
of the state is
found in the center muffin-tin.  We see that the state $n=2$ has
97.7\% in the center MT and only 2.1\% in the next shell demonstrating
how well localized this Wannier function is.  The Wannier functions
corresponding to $n=0$, $n=5$, and $n=6$ have considerable $4s$- and
$4p$-character, and $n=5$ and $n=6$ have the smallest $3d$-character.  
But they are very well
localized as well, having at least 88\% of their total weight already
within the central muffin-tin sphere.  On the other hand, all Wannier
states are mixed with respect to their $l$-character, since the
minimization procedure mixes all the $l$ characters.

\begin{figure}
\epsfxsize=3.0 truein
\centerline{\epsfbox{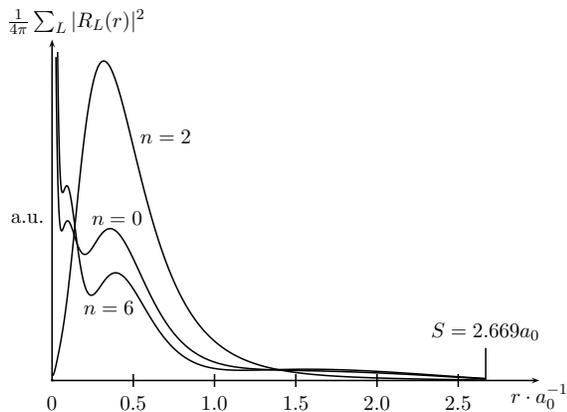}}
\caption{Radial averaged Wannier functions in the center muffin-tin sphere
  for various indices $n$.}
\label{fig:aw}
\end{figure}

Figure~\ref{fig:aw} shows a few radial averaged Wannier functions in
their center MT. We should note that the peak of the states $n$ = 0
and 6 for $r \rightarrow 0$ does not contribute very much to matrix
elements because of the $r^2$ in Eq.~(\ref{eq:I3}).
Figure~\ref{fig:aw} may also be qualitatively compared with the
Wannier function of Cu in Ref.~\onlinecite{E89}.

\begin{figure}
\epsfxsize=3.0 truein
\centerline{\epsfbox{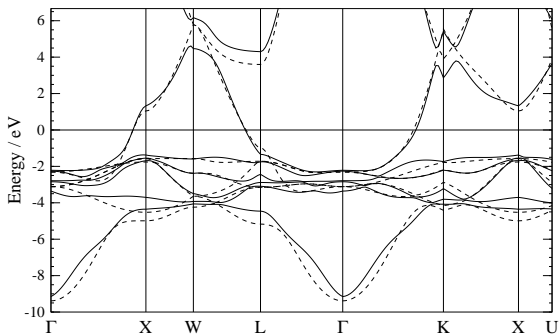}}
\caption{Comparison of LDA band structure (dashed line) and the
  diagonalized eigenvalues of Eq.~(\ref{eq:Hk}), where 3 shells
  in the lattice sum were included. The bands are relative to
  the Fermi level at 0eV.}
\label{fig:t3}
\end{figure}

From these Wannier functions we have calculated the hopping matrix
elements $t_{{\bf R}nm}$ according to Eq.~(\ref{eq:tasa}).  These can
be inserted into Eq.~(\ref{eq:Hk}) in order to determine an effective
orthogonalized (diagonal S-matrix) tight-binding representation.  The
matrices $H^{\bf k}$ are not diagonal because the unitary matrix which
was used in the minimization of the Wannier functions scrambled the
different bands.  However, we can still diagonalize $H^{\bf k}$ for
each $k$-point and compare the eigenvalues with the original LDA band
structure.  The results are shown in Fig.~\ref{fig:t3}, where we have
cutoff the ${\bf R}$-sum in Eq.~(\ref{eq:Hk}) to include only ${\bf
0}$ and the 3 nearest shells, i.e., 43 sites.  We have found that the
decay of $t_{{\bf R}nm}$ as a function of $|{\bf R}|$ is a lot faster
than that of the Fourier components of the band structure,
Eq.~(\ref{eq:tEnk}).  If we just take Eq.~(\ref{eq:tEnk}) and
recalculate the band structure according to Eq.~(\ref{eq:Hk}), the
agreement is a lot worse (for the same number of sites in $t_{{\bf
R}nn}$).  This can be understood in the following way: Labeling the
bands according to $E_n({\bf k})<E_{n+1}({\bf k})$ is not ``natural'',
therefore at points in $k$-space where two bands cross each other
$E_n({\bf k})$ has a kink.  Those kinks have non-negligible Fourier
components with large $|{\bf R}|$, which our cutoff sets to zero.

\begin{figure}
\epsfxsize=3.0 truein
\centerline{\epsfbox{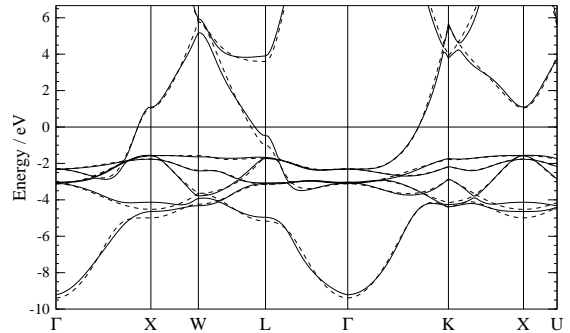}}
\caption{Comparison of LDA band structure (dashed line) and the
  diagonalized eigenvalues of Eq.~(\ref{eq:Hk}), where 8 shells
  in the lattice sum were included.}
\label{fig:t8}
\end{figure}

In Fig.~\ref{fig:t8} we have included 8 shells in the lattice sum of
Eq.~(\ref{eq:Hk}), i.e., 141 sites. As we can see the two curves agree
even better.  These calculations are a test of the quality of the
Wannier functions, i.e., how well the matrix elements obtained from
these Wannier functions reproduce the known band structure.

With respect to the magnitude of the hopping matrix elements, the
$t_{{\bf 0}nm}$ are largest and provide information about the
positions of the bands.  For a next neighbor ${\bf R}$ the $|t_{{\bf
R}nm}|$ are of the order of 0.3eV for d-states (and 1eV for the state
with $n=0$).  For larger ${\bf R}$ the hopping matrix elements are
less than 0.15eV for d-states (and less than 0.5eV for the state with
$n=0$).

\begin{figure}
\epsfxsize=3.0 truein
\centerline{\epsfbox{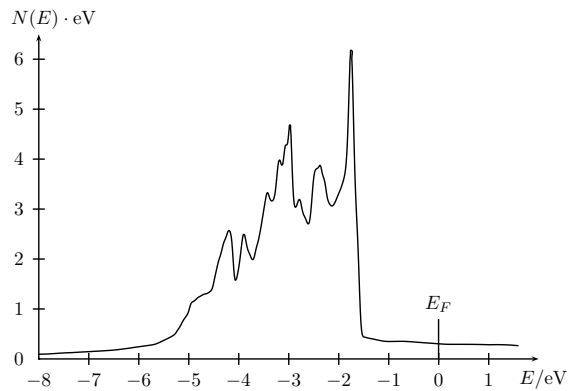}}
\caption{Total electronic density of states, relative to
  the Fermi energy.  We have used a Gaussian broadening of
  2~mRyd to remove the
  spikes inherent in the linear tetrahedron method.}
\label{fig:ds}
\end{figure}

Next, consider the projected DOS (PDOS).  In these calculations we
have used the tetrahedron method\cite{Tetra71} (see also
Ref.~\onlinecite{HS84}) with 200 ${\bf k}$-points and 691 tetrahedras
in the irreducible part of the Brillouin zone.  In Fig.~\ref{fig:ds}
we have plotted the total DOS which can also be found in the
literature (see Ref.~\onlinecite{MJW78}).  
\begin{figure}
\epsfxsize=3.0 truein
\centerline{\epsfbox{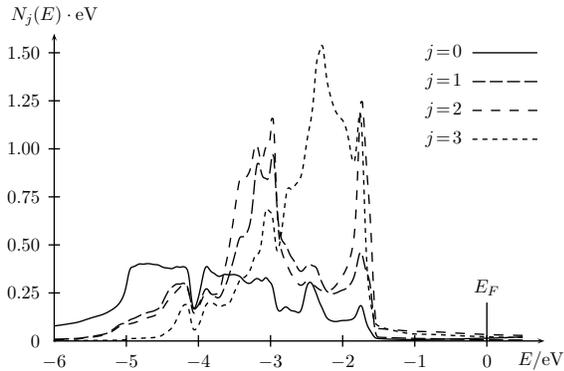}}
\caption{Projected DOS for Wannier states 0 through 3.}
\label{fig:w0}
\end{figure}

\begin{figure}
\epsfxsize=3.0 truein
\centerline{\epsfbox{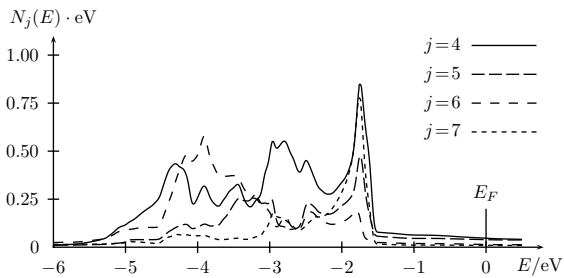}}
\caption{Projected DOS for Wannier states 4 through 7.}
\label{fig:w4}
\end{figure}

Figures~\ref{fig:w0} and
\ref{fig:w4} show the projected DOS according to Eq.~(\ref{eq:wDOS})
for the Wannier functions 0 through~7. It is interesting that states
$j=2$ and 3 have very similar $\ell$ character
(cf.~Table~\ref{tab:lcomp}) but a very different projected DOS in
Fig.~\ref{fig:w0}, i.e., they are peaked at different energies, and
emphasize different parts of the $d$ band.
Table~\ref{tab:decomp-nosEF} shows the projected density of states
(actually the percentage of the DOS in different states) and the
projected number of states evaluated at the Fermi level, where the
$j$th projected number of states is defined as
\begin{equation}
  \nonumber
  n_j(E_F) = \int^{E_F} dE ~ N_j(E) ~.
\end{equation}
This is just the number of electrons in the $j$th state. Every state could
maximally be occupied with 2 electrons (one for each spin direction).
\parbox{8cm}{
\begin{table}
  \caption{Projected number of states at the Fermi energy, i.e., $n_j(E_F)$.
    For $j>8$ (the numbers not given) $n_j(E_F)<0.25$.
    The second line shows the percentage of the DOS at Fermi energy,
    i.e., $100\cdot N_j(E_F)/N(E_F)$.}
  \label{tab:decomp-nosEF}
  \begin{tabular}{c|ccccccccc}
  $j$        &    0 &    1 &    2 &    3 &    4 &    5 &   6 &   7 &    8 \\
\hline
  $n_j(E_F)$ & 1.17 & 1.31 & 1.59 & 1.80 & 1.34 &  .66 & .90 & .51 &  .78 \\
  \% DOS     &  4.7 &  2.1 & 11.1 &  7.1 & 15.4 & 13.0 & 4.6 & 3.2 & 27.3 \\
  \end{tabular}
\end{table}}

We next consider a calculation of the direct Coulomb integral $U_{ii}$
for a $d$-like orbital with itself.  As discussed above, the Wannier
function for $n=2$ has nearly perfect $d$-character and to a good
approximation we can consider only its contribution in the central MT,
i.e., at site ${\bf 0}$.  What we then calculate is the on-site
Coulomb matrix element between two electrons (with different spin
because of the Pauli principle) at the same site in the same Wannier
state, i.e., essentially the Hubbard-U in its original
sense.\cite{H63} The method described in Sec.~\ref{sec:met1} yields
$U(dd;{\bf 0},{\bf 0})=$25.26 eV while the method from
Sec.~\ref{sec:fft} yields 25.16 eV for this quantity.  But with the
second method we are able to calculate all the elements involving
tails of the Wannier functions in other muffin-tins in the double sum
in Eq.~(\ref{eq:W1234a}).  When we do this and include sites where
${\bf R}_i$ and ${\bf R}_j$ are nearest neighbors, we get
$U_{dd}=$25.51eV, which shows that including the portions of the
Wannier function on neighboring sites is a rather small correction on
$U$ for such a strongly localized function.  Table~\ref{tab:onsiteFFT}
show these quantities for the Wannier functions $n=0$ through~6.
Going beyond nearest neighbors would have even a smaller effect.
Therefore, one can truncate the sums over higher neighbor shells for
the Coulomb matrix elements, which converge faster than for the
hopping matrix elements.  The reason for this is that the Coulomb
integral involves a product of four wave functions, whereas the
hopping matrix elements involve only two wave functions.
\parbox{7cm}{
\begin{table}
  \caption{Onsite FFT $U$'s. In the first line (onsite-$U$)
    we have only included
    ${\bf R}_i={\bf R}_i={\bf 0}$ in Eq.~(\ref{eq:W1234a}), i.e.,
    $U(jj;{\bf 0},{\bf 0})$.  The second line (NN-$U$)
    shows the same quantity, where we have
    included nearest neighbors for ${\bf R}_i$ and ${\bf R}_j$.}
  \label{tab:onsiteFFT}
  \begin{tabular}{c|cccccccc}
  $j$        &     0 &     1 &     2 &     3 &     4 &     5 &     6 \\
\hline
  onsite-$U$ & 14.82 & 19.05 & 25.16 & 25.49 & 20.79 & 13.81 & 13.22 \\
  NN-$U$     & 16.29 & 19.81 & 25.51 & 25.86 & 21.44 & 15.32 & 14.97 \\
  \end{tabular}
\end{table}}

The FFT approach allows us to calculate Coulomb matrix elements for
Wannier functions centered on different sites. We have done this for
the states $a=\{ {\bf 0}, 2 \}$ and $b=\{ {\bf R}, 2 \}$ where ${\bf
R}$ is a nearest neighbor of ${\bf 0}$. Both are $d$-like states. In
Eq.~(\ref{eq:W1234a}) we have again included nearest neighbors for
${\bf R}_i$ and ${\bf R}_j$. The result is $U_{ab}$=5.87 eV.  The
largest contribution in the sum is $U(ab;{\bf 0},{\bf R})=$5.66 eV,
which is the contribution arising from the two center spheres of
states $a$ and $b$.

Our first method is most useful for calculating inter-band (on-site)
Coulomb matrix elements when the states are so well localized that we
can neglect the contribution from neighboring spheres.  We have
calculated both the direct Coulomb matrix elements $U_{nm}$ and the
exchange integrals $J_{nm}$ for all $n$ and $m$.  Here the $n,m$ just
indicates the band and all Wannier states are centered at site ${\bf
0}$.  The results are given in Tables~\ref{tab:Unm} and \ref{tab:Jnm}
for the first 7~bands.  We see that the on-site intra-band Coulomb
matrix elements are largest for the Wannier states $n$=2 and 3, which
have almost pure d-character.  We also note that all the direct
Coulomb matrix elements $U_{nm}$ are rather large, while the exchange
Coulomb matrix elements $J_{nm}$ with $n \ne m$ are rather small (note
that the diagonal terms for both $U_{ii}$ and $J_{ii}$ are identical
by definition, cf.\ Eq.~(\ref{eq:UJ})).  When we compare the diagonal
elements $U_{nn}(=J_{nn})$ from Tables \ref{tab:Unm} and \ref{tab:Jnm}
with the first line of Table~\ref{tab:onsiteFFT} we note relatively
large differences for the states $n$=0, 5, and 6, which have large $s$
and $p$ character, as can be seen from Table~\ref{tab:lcomp} (e.g. 0.52eV for
$n$=0).  This leads to a peak in their charge density near ${\bf r=0}$ as we
can see from Fig.~\ref{fig:aw} for $n$=0 and~6.  For those states our FFT
calculations had a numerical problem because our real-space grid was
too large (with $\Delta x=0.17\, a_0$).  But for $n$=2, we do not have
a peak at ${\bf r=0}$, and the FFT approach does an excellent job.
\parbox{7cm}{
\begin{table}
  \caption{Inter-band on-site matrix elements: $U_{nm}$ in eV.}
  \label{tab:Unm}
  \begin{tabular}{c|ccccccc}
        & $m=0$ &     1 &     2 &     3 &     4 &     5 &     6 \\ \hline
  $n=0$ & 14.30 & 15.86 & 17.86 & 17.45 & 16.31 & 13.24 & 12.77 \\
      1 & 15.86 & 19.02 & 21.36 & 20.98 & 19.33 & 15.25 & 14.42 \\
      2 & 17.86 & 21.36 & 25.26 & 24.13 & 22.30 & 17.09 & 15.86 \\
      3 & 17.45 & 20.98 & 24.13 & 25.26 & 21.88 & 16.69 & 16.12 \\
      4 & 16.31 & 19.33 & 22.30 & 21.88 & 20.70 & 15.76 & 14.78 \\
      5 & 13.24 & 15.25 & 17.09 & 16.69 & 15.76 & 13.23 & 12.32 \\
      6 & 12.77 & 14.42 & 15.86 & 16.12 & 14.78 & 12.32 & 12.43 \\
  \end{tabular}
\end{table}

\begin{table}
  \caption{Inter-band on-site matrix elements: $J_{nm}$ in eV.}
  \label{tab:Jnm}
  \begin{tabular}{c|ccccccc}
        & $m=0$ &     1 &     2 &     3 &     4 &     5 &     6 \\ \hline
  $n=0$ & 14.30 &  0.91 &  0.73 &  0.76 &  0.92 &  0.98 &  1.34 \\
      1 &  0.91 & 19.02 &  1.22 &  0.84 &  0.91 &  1.43 &  0.69 \\
      2 &  0.73 &  1.22 & 25.26 &  0.92 &  1.14 &  0.95 &  0.58 \\
      3 &  0.76 &  0.84 &  0.92 & 25.26 &  0.99 &  0.71 &  0.62 \\
      4 &  0.92 &  0.91 &  1.14 &  0.99 & 20.70 &  1.20 &  0.79 \\
      5 &  0.98 &  1.43 &  0.95 &  0.71 &  1.20 & 13.23 &  1.22 \\
      6 &  1.34 &  0.69 &  0.58 &  0.62 &  0.79 &  1.22 & 12.43 \\
  \end{tabular}
\end{table}}

The Hubbard-$U$ clearly depends on the specific shape of the Wannier
functions.  Intuitively, one expects bigger $U$'s for more localized
orbitals.  As an example of this, we have calculated a less localized
Wannier state (by doing fewer steps in the minimization procedure).
In this case, the highest $d$-character state, which is almost pure
$d$-like, has only 58.5\% of its charge density in the center
muffin-tin, and 95.4\% within the first 3 shells.  For those three
shells we have used the FFT method to calculate all $43^2$ terms
(\ref{eq:W1234b}).  We find a $U=13.8$eV for this less localized
$d$-state.

We should also note that most model calculations assume very
localized, pure (in $l$-character) Wannier functions.  In particular,
they often assume that LDA or some one-electron-like treatment is
adequate for non-$d$ and non-$f$ electron states, and that the only
explicit correlations that need to be included are related to onsite
(or sometimes also nearest-neighbor) Coulomb $U$'s for the $d$ (or
$f$) states.  It is also often implicitly assumed that the non-$d$ and
non-$f$ states have some screening contribution to the effective $U$'s
in the model Hamiltonian.  These types of assumptions raise some
difficulties for us to connect our treatment to the model
Hamiltonians, since the orthogonalization properties and mixing
necessary for localizing our Wannier functions scramble the
$l$-character of the resulting orbitals.  Hence, our effective $U$'s do
not have a pure $d$ or $f$ character (or $s$ or $p$).  Also, since we
calculate $U$'s for all of the orbitals, we are implicitly including
correlation effects for all orbital ($s$ and $p$ as well as $d$ and
$f$), and however the $U$'s in our treatment are ultimately screened
in some many-body treatment, this screening may be different from that
assumed in the model Hamiltonians.  We may ultimately be forced to use
some kind of projection of our orbitals onto pure $l$-character states
in order to make appropriate identification between our types of
states and more conventional model Hamiltonians.

\section{Constrained LDA}
\label{sec:constrain}

\begin{figure}
\epsfxsize=3.0 truein
\centerline{\epsfbox{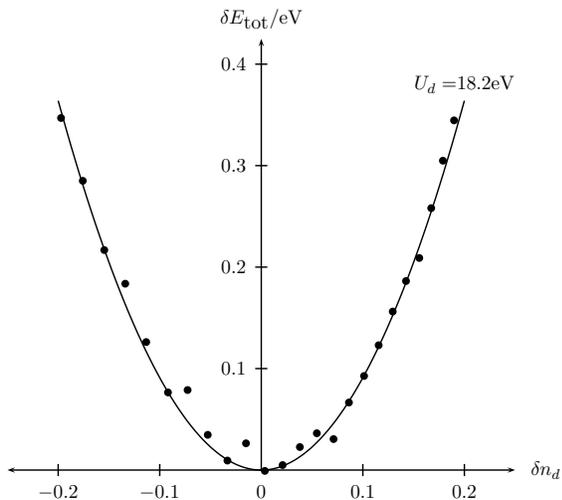}}
\caption{The total energy as a function of the
  effective change in $d$ charge.  The line is a quadratic fit.}
\label{fig:cl}
\end{figure}

Finally, we have done a constrained LDA calculation\cite{DBZA84} to
obtain an estimate for the Hubbard $U$.  In this method the Hubbard
$U$ is defined as the Coulomb energy cost to place two (in our case
$d$) electrons at the same site.  This is
\begin{equation}
  \label{eq:clda.1}
  U = E(N_d+1) + E(N_d-1) - 2 E(N_d) ~.
\end{equation}
Here $E(N_d)$ is the ground-state energy with $N_d$ $d$-electrons.  If
we consider this energy as a continuous function of $N_d$, where we
constrain the value of $N_d$ to be away from its minimized value, then
the Hubbard $U$ is given by:
\begin{equation}
  \label{eq:clda.2}
  U_{dd} = \frac{\delta^2 E(N_d)}{\delta N_d^2}
\end{equation}
This constraint, which fixes the total number of d-electrons to be
$N_d$, can be taken into account by adding a Lagrange parameter $v_d$
to the total energy; i.e., the energy of the constrained system is
given by
\begin{equation}
  \label{eq:clda.3}
  E(N_d) = \mbox{min}[E\{n({\bf r})\}+v_d\{ 
  \int d^3r n_d({\bf r}) - N_d \}]~.
\end{equation}
Here $E\{n({\bf r})\}$ is the usual band-structure energy and
$n_d({\bf r})$ is the $d$-electron density.  On minimization the extra
term in Eq.~(\ref{eq:clda.3}) leads to an additional constant
potential, $v_d$, in the Kohn-Sham equations, which acts only on the
$l=2$ angular momentum components of the wave function.  Within the
LMTO method, this is accomplished by adding a constant potential,
$v_d$, when solving the radial Schr{\"o}dinger equation for $l=2$, and
then calculating the total energy as a function of $v_d$.  Since each
value of $v_d$ changes the $d$ occupation number, the final result can
be written as $E(N_d)$.  This dependence is shown in Fig.~\ref{fig:cl}
and can be accurately fitted by a parabola, $\delta
E=\frac{1}{2}U_{dd}N_d^2$, with $U_{dd}=18.2$ eV. This is of the same
magnitude as the result obtained from the direct calculation of the
Coulomb matrix elements, even though one might expect a smaller value
because of the screening effects that are believed to be included in
this calculation.  In our calculations we have only used a one-atom
unit cell.  If a larger unit cell is chosen, one could do a variety of
additional constraints (e.g., changing the $d$-occupation separately
on two different atoms).  Such calculations could attempt to sort out
more details of effective Hamiltonians (perhaps even two-particle
parameters).  However, such calculations would take our work in a
different direction from what we are interested.  Also, given the
intuitive nature of the constrained method and the difficulties in
fitting such a large parameter space, it is not clear how useful the
resulting parameters would be or their uniqueness.

\section{Conclusions}
\label{conc}

We have shown in this paper that ab-initio band-structure methods can
be used for a first-principles calculation of well localized Wannier
functions, which is achieved by using a method proposed by Marzari and
Vanderbilt.\cite{MV97} From these localized Wannier functions the
on-site and inter-site one-particle matrix elements of the Hamiltonian
can be calculated.  A good localization of the Wannier functions is
needed to keep tight-binding (hopping) matrix elements restricted to a
small number of near neighbors.  The Coulomb matrix elements within
these localized Wannier states can also be calculated and are
similarly only important between on-site and nearest-neighbor Wannier
functions.  The result is thus an electronic multi band Hamiltonian in
second quantization with first-principles one- and two-particle matrix
elements.  The Hamiltonian is of the form of an extended multi-band
Hubbard model but without adjustable parameters; the parameters are
directly calculated for a given material.  The only approximations
still involved are the ones inherent to the ab-initio band structure
method used (e.g., the muffin-tin assumption, the ASA approximation,
the choice of linearized orbitals in the LMTO, and the ''frozen-core''
approximation), and the truncation in the number of bands (states) per
site that is explicitly taken into account (truncation of the $l$-sum).
The resulting multi-band Hamiltonian that includes the Hubbard-$U$
terms, of course, still has to be studied within a reliable many-body
method or approximation, e.g, a multi-band version of the DMFT as in
Refs.~\onlinecite{APKAK97}, \onlinecite{KL99}, \onlinecite{ZPKPNA00},
and \onlinecite{NHBPAV00}.

Our Cu calculations yield on-site direct Coulomb matrix elements
(''Hubbard-U's'') of the magnitude of 20 eV for 3d-Wannier-states and
inter-site (Hubbard-U's between nearest-neighbors) values of 5 eV.
This is the magnitude discussed already earlier\cite{H63} and similar
to those for atomic 3d-states.  These U-values are much larger than
commonly expected or used in model studies.  Although our calculated
Coulomb matrix elements are unscreened, the constrained LDA, which
includes some screening effects, gives comparable magnitudes for U.
Since dynamic screening due to the mobile electrons in the outer
shells (bands) is taken into account automatically when using an
appropriate many-body method, e.g., a generalized random phase
approximation (RPA), the only screening that should be included in a
better theory is a static, short range screening by the inner core
electrons.  However, the (atomic like) electronic states representing
the inner (''frozen'') core are well known, and it should be possible
to calculate their screening contribution from a (generalized) static
Lindhard theory.  In future work we plan to examine the static
screening due to the inner core states, an application of appropriate
(multi-band) many-body methods, and the application to more strongly
correlated 3-d materials such as iron, cobalt, and nickel.  Any
treatment of screening will, of course, have to be very careful that
screening effects are not double counted (once in the explicit
screening and then a second time when the many-body Hamiltonian is
solved).  Finally, although any localized orbitals could be used as
the basis for a many-body treatment, the approach we have used (of
constructing localized orbitals from LDA band states) has the
advantage that these orbitals are a good basis set for any states
without strong electron-electron correlations, since LDA is believed
to be an accurate approximation in this limit.  We can hope that an
additional more explicit treatment of the strong correlations by a
many-body theory will correct and improve on the LDA starting point.

\acknowledgements

This research was partially supported by the Department of Energy
under contract W-7405-ENG-36.  This research used resources of the
National Energy Research Scientific Computing Center, which is
supported by the Office of Science of the U.S. Department of Energy
under Contract No. DE-AC03-76SF00098.  We also acknowledge a generous
grant from the University of Bremen for a visit by one of us (RCA)
that helped make possible this work.  We thank Veljko Zlatic for
suggesting the application of the FFT algorithm in the Coulomb matrix
element calculations.

\appendix

\section{Spherical harmonics expansions}
\label{app:spherical}

Any function $A({\bf r})$ within a (muffin-tin) sphere may be expanded
in terms of spherical harmonics:
\begin{equation}
  \label{eq:SH1}
  A({\bf r}) = \sum_L A_L(r) Y_L(\hat{\bf r})
\end{equation}
If two functions $A({\bf r})$ and $B({\bf r})$ are given via their
coefficients $A_L(r)$ and $B_L(r)$, then the corresponding
coefficients $F_L(r)$ of the function $F({\bf r})=A({\bf r})B({\bf
r})$ are given by:
\begin{equation}
  \label{eq:SH2}
  F_L(r) = \sum_{L_1,L_2} A_{L_1}(r) B_{L_2}(r) C_{L_1LL_2}
\end{equation}
The Gaunt coefficients $C_{LL'L''}$ are defined by
\begin{eqnarray}
  \label{eq:gaunt}
  C_{LL'L''} &=& \int d^2\Omega \; 
  Y_L(\Omega) \; Y^*_{L'}(\Omega) \; Y_{L''}(\Omega)
\nonumber \\
  &=& \delta_{m'',m'-m} ~ \sqrt{ \frac{2\l''+1}{4\pi} } ~ c^{l''}(L',L)
\end{eqnarray}
and the $c^k(L',L)$ are tabulated in Ref.~\onlinecite{CS53}.  We may
use Eq.~(\ref{eq:SH2}) to multiply a function with a plane wave
$e^{\bf k}({\bf r}) \equiv e^{-i{\bf kr}}$ whose coefficients are
given by (see Ref.~\onlinecite{JDJ75}):
\begin{equation}
  \label{eq:SH3}
  e^{\bf k}_L(r) = 4\pi j_l(kr) \left[ i^l Y_L(\hat{\bf k}) \right]^*
\end{equation}


%
%

%
%

\end{document}